\documentclass[12pt]{iopart}

\usepackage{iopams} 
\usepackage{cite}
\usepackage{graphicx} 
\begin{document}

\title[Soft-boundary graphene nanoribbon]{Soft-boundary graphene nanoribbon formed by a graphene sheet above
a perturbed ground plane: conductivity profile and SPP modal current
distribution}

\author{Ebrahim Forati and George W. Hanson}

\address{Department of Electrical Engineering and Computer Science, University of Wisconsin Milwaukee, WI53211, USA}
\ead{\mailto{eforati@uwm.edu},\mailto{george@uwm.edu}}
\begin{abstract}
An infinite sheet of graphene lying above a perturbed ground plane
is studied. The perturbation is a two dimensional ridge, and a bias
voltage is applied between the graphene and the ground plane, resulting
in a graphene nanoribbon-like structure with a soft-boundary (SB)
The spatial distribution of the graphene conductivity forming the
soft-boundary is studied as a function of the ridge parameters and
the bias voltage. The current distribution of the fundamental TM surface
plasmon polariton (SPP) is considered. The effect of the ridge parameters
and shape of the soft boundary on the current distributions are investigated,
and the conditions are studied under which the mode remains confined
to the vicinity of the ridge region.

\end{abstract}

\maketitle
\section{Introduction\label{sec:Introduction} }

Graphene is a two-dimensional material having unique electronic, mechanical,
and optical properties \cite{Novoselov,Zhang,Berger,Geim,Nair,Bonaccorso}. A variety
of applications have been considered, including optical sensors \cite{Schedin},
transparent electrodes, nanoelectromechanical applications (NEMs)
\cite{Geim_science}, and optoelectronic applications \cite{Mak,Mueller,Xia,Lee}.
Graphene's interesting properties are partly because of its conical
conduction and valance bands joined by two points at the Fermi level
\cite{Castro}. Graphene, doped with excess carriers, can also guide
surface plasmon oscillations at terahertz frequencies, similar to
those in noble metals at infrared frequencies \cite{Johan-Christensen,GWHanso}.
In this regard, graphene is considered a better plasmonic material than nobel metals with greater confinement of the electromagnetic energy and lower loss. Compared to noble-metal plasmons, graphene modes have two major advantages, 1) they are long-lived excitations because of the low loss of graphene, and 2) their frequency can are controlled by electrostatic doping. The tunability of transverse magnetic (TM) surface plasmons is due to the ability to vary the carrier density, which can be easily
achieved by gate biasing or chemical doping. Graphene can also support transverse electric (TE) surface plasmons which are loosely confined to its surface; we do not consider them further in this paper.
Electron energy-loss spectroscopy (EELS) was first used to prove the existence of the plasmonic effect in graphene experimentally  \cite{N19,N21}. Later, surface plasmons were excited with optical means and the interaction of optical phenomena with graphene plasmons was  studied experimentally \cite{N22,N26}.
Considering only TM surface plasmons, an infinite suspended sheet of graphene supports one surface mode. However, a graphene strip supports an infinite number of 2D-bulk modes and two almost degenerate symmetrical and anti-symmetrical edge modes. Therefor, graphene strips are of obvious interest for waveguiding and related applications due to the variety of possible modes that may propagate. Also, plasmons in graphene with a magnetic field present have been studied \cite{N51,N53} and shown to have interesting properties. For example, a magnetically biased graphene strip supports edge and bulk magnetoplasmons with nonreciprocal properties.  
 Electrically, graphene can be modeled by a surface conductivity which is considered
in several works \cite{Falkovsky,Falkovsky-1,Mikhailov,GusyninP,Gusynin-1,Peres,W-Hanson,Hanson-1,Ziegler}. Here we use the
local conductivity resulting from the Kubo formula  \cite{Gusynin}

\[
\sigma\left(x\right)=\frac{je^{2}}{\pi\hbar^{2}\left(\omega-j\Gamma\right)}\intop_{0}^{\infty}\varepsilon\left(\frac{\partial f_{d}\left(\varepsilon,x\right)}{\partial\varepsilon}-\frac{\partial f_{d}\left(-\varepsilon,x\right)}{\partial\varepsilon}\right)d\varepsilon\qquad\qquad\qquad\qquad
\]

\begin{equation}
\qquad\qquad\,-\frac{je^{2}\left(\omega-j\Gamma\right)}{\pi\hbar^{2}}\intop_{0}^{\infty}\frac{f_{d}\left(-\varepsilon,x\right)-f_{d}\left(\varepsilon,x\right)}{\left(\omega-j\Gamma\right)^{2}-4\left(\varepsilon/\hbar\right)^{2}}d\varepsilon,\label{eq:cond}
\end{equation}
where $-e$ is the charge of an electron, $\hbar$ is the reduced
Plank's constant, $f_{d}\left(\varepsilon,x\right)=\left(exp\left(\frac{\varepsilon-\mu_{c}\left(x\right)}{k_{B}T}\right)+1\right)^{-1}$is
the Fermi-Dirac distribution, $k_{B}$ is the Boltzmann's constant,
$\mu_{c}\left(x\right)$ is the inhomogeneous chemical potential created
by the bias, and $\Gamma=10^{13}$ 1/s is the phenomenological scattering
rate. 

\begin{figure}[t]
\begin{centering}
\includegraphics[scale=0.4]{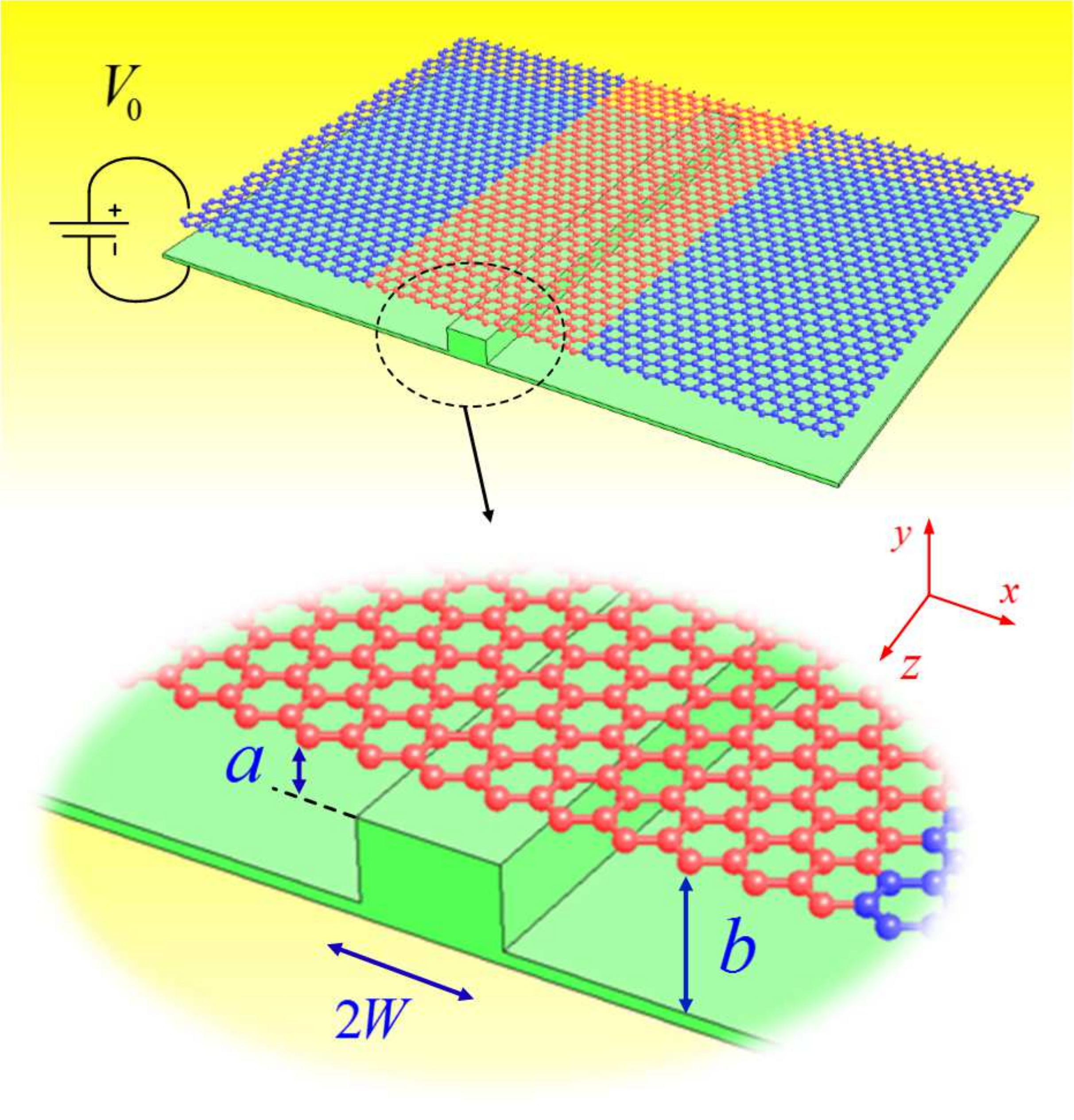}
\par\end{centering}

\caption{Electrostatically-biased graphene sheet over a ground plane with a
two dimensional ridge. }
\end{figure}

The first term in (\ref{eq:cond}) is due to intraband contributions
and the second term is due to interband contributions. The sign of
$\mathrm{Im}\left(\sigma\right)$ is negative and positive for the
intraband and interband contributions, respectively. Therefore, depending
on the parameters in (\ref{eq:cond}), such as frequency and temperature,
one of the two contributions dominates and determines the sign of
$\mathrm{Im}\left(\sigma\right)$. 

It can be shown that TM surface waves can propagate only if $\mathrm{Im}\left(\sigma\right)<0$
\cite{Mikhailov,W-Hanson}. This phenomena is exploited in \cite{Vakil},
where it is suggested that a graphene sheet and an inhomogeneous biasing
scheme, such as that resulting from a ground plane ridge (Fig. 1),
can be used to electronically form a conductivity profile capable
of confining SPP propagation. That is, in \cite{Vakil} the ridge
is assumed to achieve a piece-wise constant conductivity profile with
$\mathrm{Im}\left(\sigma\right)<0$ in the desired channel region
$\left|x\right|<W$ and $\mathrm{Im}\left(\sigma\right)>0$ outside
of the channel, $\left|x\right|>W$, forming, essentially, a hard-boundary
(HB) graphene nanoribbon (GNR). In this paper, we investigate this
structure (Fig. 1) without the piece-wise constant conductivity assumption
- the biased ridge/ground plane results in an electrostatic (bias)
charge distribution $\rho\left(x\right)$ determined from Laplace's
equation, which, in turn, results in the inhomogeneous chemical potential
$\mu_{c}\left(x\right)$ such that $\sigma=\sigma\left(x\right)$.
This geometry allows the ability to tune $\mathrm{Im}\left(\sigma\right)$
to be negative in a limited area (in the vicinity of the ridge), forming
a channel for SPP guiding, albeit forming a soft boundary. In particular,
it is impossible to form the hard boundary case using the ridged ground
plane, but one can approximate the HB case with a sufficiently-sharp
soft boundary, as shown below.

The time convention is $e^{j\omega t}$ and the temperature in (\ref{eq:cond})
is set to be $T=3$ K, consistent with \cite{Vakil}, since at lower
temperature the interband contribution can dominate the intraband
contribution down to lower frequencies then at room temperature. For
example, at $f=45\:\mathrm{THz}$ and $\mu_{c}=0.05\:\mathrm{eV},$
the intraband and interband contributions at $T=3$ K are $\sigma_{\mathrm{intra}}=1.4-j41\:\mu\mathrm{S}$
and $\sigma_{\mathrm{inter}}=8.9+j62\:\mu\mathrm{S}$ while at $T=300$
K they are $\sigma_{\mathrm{intra}}=1.4-j42\:\mu\mathrm{S}$ and $\sigma_{\mathrm{inter}}=27+j39\:\mu\mathrm{S}$. 

In the following, properties of the resulting channel are studied
as a function of the parameters shown in figure 1. Then, the current
distribution of the fundamental mode of the geometry is considered
and the conditions are explored under which the mode will remain confined
to the vicinity of the step region. One interesting result is that
currents can still be concentrated to the vicinity of the ridge even
when $\mathrm{Im}\left(\sigma\right)$ is negative everywhere. This
requires some special conditions which are discussed toward the end
of this work.

\section{Methodology and formulations}

\begin{figure}[tbh]
\begin{centering}
\includegraphics[scale=0.4]{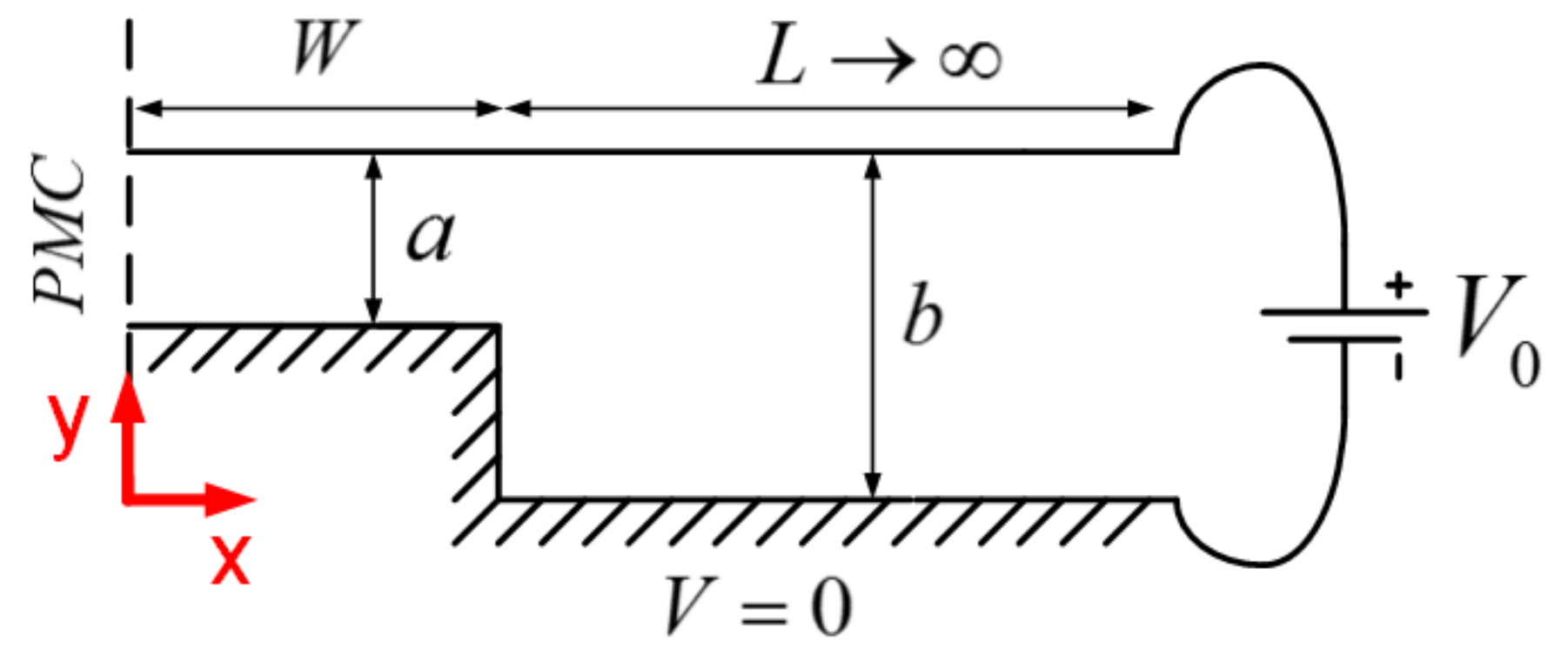}
\par\end{centering}

\centering{}\caption{The x-y view of the geometry of figure 1.}
\end{figure}

Figure 2 shows the x-y view of the geometry in figure 1. A perfect
magnetic conductor sheet (PMC) is placed at $x=0$ since the geometry
is symmetrical with respect to $x=0$. By solving Laplace's equation
and applying the appropriate boundary conditions, it is easy to show
that the bias voltage distribution between the graphene and ground
plane is

\begin{equation}
\frac{V(x,y)}{V_{0}}=\left\{
\begin{array}
[c]{ll}%
{1+\frac{y-b}{a}} & \quad \left|x\right|<W\\
\frac{y}{b}+{\displaystyle \sum_{n=1}^{\infty}}C_{n}sin\left(\frac{n\pi}{b}y\right)e^{-\frac{n\pi}{b}\left(\left|x\right|-W\right)}
 & \quad  \left|x\right|>W
\end{array}
\label{eq:V}%
\right.
\end{equation}

\noindent where 
\begin{equation}
C_{n}=-\frac{2b}{a}\left(\frac{1}{n\pi}\right)^{2}sin\left(n\pi\left(1-\frac{a}{b}\right)\right).\label{eq:Cn}
\end{equation}

In obtaining (2),  a zeroth order approximation has been used to assume an x- independent potential in the region above the step ($\left|x\right|<W$). Otherwise, the problem needs to be solved numerically (e.g., by expanding the potentials as series for both $\left|x\right|<W$ and $\left|x\right|>W$ regions). The zeroth-order solution is a good approximation for $W\ll b$ and/or $a\ll b$ in Fig. 1.

Therefore, the electrostatic surface charge density on the graphene
sheet is

\begin{equation}
\frac{\rho\left(x\right)}{\varepsilon_{0}V_{0}}=\left\{
\begin{array}
[c]{ll}%
{\frac{1}{a}} & \quad \left|x\right|<W\\
\frac{1}{b}+{\displaystyle \sum_{n=1}^{\infty}}\frac{n\pi}{b}C_{n}\left(-1\right)^{n}e^{-\frac{n\pi}{b}\left(\left|x\right|-W\right)}
 & \quad  \left|x\right|>W
\end{array}
\label{eq:rho}%
\right.
\end{equation}

\noindent which can be used to find the chemical potential on the graphene sheet
as

\begin{equation}
\mu_{c}\left(x\right)=\frac{\hbar}{e}v_{F}\sqrt{\frac{\pi\rho\left(x\right)}{e}}\label{eq:chemical pot}
\end{equation}
where $v_{F}=9.546\times10^{5}$ m/s is the Fermi velocity. Eq. (\ref{eq:cond})
then gives the conductivity distribution $\sigma\left(x\right)$ on
the graphene sheet. 

In order to find the dynamic modal current distributions on the graphene
(eigencurrents of the structure), Ohm's law can be used in the one
dimensional Fourier transform domain $z\leftrightarrow\beta_{z}$
as

\begin{equation}
\mathbf{J}\left(x,\beta_{z}\right)=\sigma\left(x\right)\mathbf{E}\left(x,b,\beta_{z}\right),\label{eq:Ohms}
\end{equation}
where the Fourier transform pair is defined as

\begin{equation}
\mathbf{E}\left(x,y,\beta_{z}\right)=\intop_{-\infty}^{\infty}\mathbf{E}\left(x,y,z\right)e^{-j\beta_{z}z}dz
\end{equation}

\begin{equation}
\mathbf{E}\left(x,y,z\right)=\frac{1}{2\pi}\intop_{-\infty}^{\infty}\mathbf{E}\left(x,y,\beta_{z}\right)e^{j\beta_{z}z}d\beta_{z}.
\end{equation}
Green's theorem relates the current and the electric field as 

\begin{equation}
\mathbf{E}\left(x,y,\beta_{z}\right)=\left(k_{0}^{2}+\nabla_{\beta_{z}}\nabla_{\beta_{z}}.\right)\qquad\qquad\qquad\label{eq:elec}
\end{equation}

\[
\qquad\qquad\qquad\intop_{x^{\prime}}g\left(x,y,x^{\prime},\beta_{z}\right)\frac{\mathbf{J}\left(x^{\prime},\beta_{z}\right)}{j\omega\varepsilon_{0}}dx^{\prime}
\]
where 
\begin{equation}
\nabla_{\beta_{z}}=\frac{d}{dx}\hat{\mathbf{x}}+\frac{d}{dy}\hat{\mathbf{y}}+j\beta_{z}\hat{\mathbf{x}}.
\end{equation}
The Green's function in (\ref{eq:elec}) is \cite{Chew-1}

\begin{equation}
g\left(x,y,x^{\prime},\beta_{z}\right)=\frac{1}{2\pi}K_{0}\left(\sqrt{\beta_{z}^{2}-k_{0}^{2}}\sqrt{\left(x-x^{\prime}\right)^{2}+\left(y-b\right)^{2}}\right),\label{eq:Greens}
\end{equation}
where $K_{0}\left(x\right)$ is the zero's order modified Bessel function
of the first kind. Since we are considering modes tightly bound to
the graphene surface, once the ridged ground plane is used to obtain
the electrostatic bias charge density we assume that the ground plane
does not interact with the tightly-confined modal fields, which we
verified to be true. 

In summary, we assume the graphene sheet forms a conductive surface,
we find the electrostatic potential distribution $V\left(x,y\right)$
via Laplace's equation, leading to the electrostatic charge distribution
and the resulting chemical potential, resulting in the conductivity
$\sigma\left(x\right)$. Equations (\ref{eq:Ohms}) and (\ref{eq:elec})
form an integral equation whose null space gives the modes of the
structure (i.e. different $\beta_{z}$ and their associated currents).

The pulse function collocation method is used to solve the integral
equation, with point matching at the center of the pulses. The conductivity
distribution based on the electrostatic charge distribution in (\ref{eq:Ohms})
is assumed to be only slightly perturbed by the modal fields, i.e.,
$\nabla{\scriptscriptstyle \bullet}\,\mathbf{J}/j\omega\ll\rho$ where
$\rho$ is the static charge density (\ref{eq:rho}) and $\mathbf{J}$
is the dynamic modal current density (\ref{eq:Ohms}). To see that
this inequality is satisfied, assume a typical frequency of $f=30$
THz and strip width $W=25\mbox{ nm}$. If the modal current is as
large as $I=\left|\mathbf{J}\right|2W=1$ mA, then the left side of
the inequality is $10^{-10}$ C/$\mathrm{m^{2}}$. Using (\ref{eq:rho})
with typical values $a=25\:\mathrm{nm}$ and $V_{0}=20\;\mathrm{V}$
leads to $\rho=7$ mC/$\mathrm{m^{2}}$ (consistent with typical doping
densities of $4\times10^{12}$ $\mathrm{cm^{-2}}$), and the inequality
is strongly satisified.

\section{Results and discussions}

\begin{figure*}[!tbp]
\begin{centering}
\includegraphics[scale=0.2]{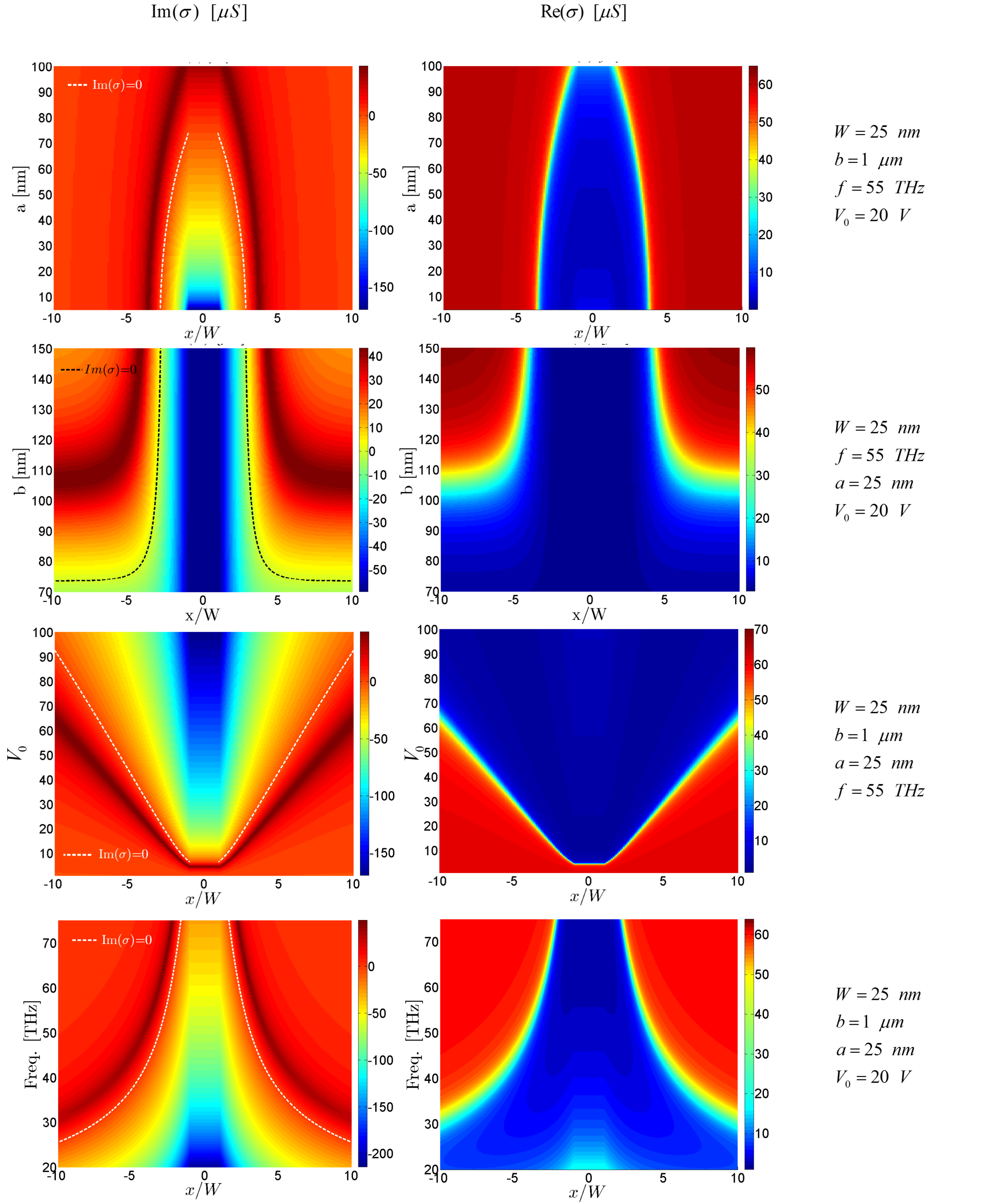}
\par\end{centering}

\centering{}\caption{Imaginary (left) and real (right) parts of the conductivity distribution
on the graphene sheet as a function of $a$, $b$, $V_{0}$, and frequency. }
\end{figure*}
Figure 3 shows the conductivity distribution for the structure of figure
1 as a function of the ridge parameters ($a$ and $b$), bias voltage, and
frequency. The dashed lines in the plots for $\mathrm{Im}\left(\sigma\right)$
specify lines where $\mathrm{Im}\left(\sigma\right)=0$, and so the
distance between the dashed lines specifies the effective width of
the channel created above the ridge with negative $\mathrm{Im}\left(\sigma\right)$.
As can be seen in figure 3, the width of the channel increases by increasing
the bias voltage, by decreasing $a$ or $b$, or by decreasing frequency
(assuming that the other parameters are fixed in each case). The width
of the channel is also more sensitive to the applied bias voltage
then the other parameters. The parameter $a$ along with the bias
voltage determines the value of the conductivity in the $\left|x\right|<W$
region. The parameter $b$ along with the bias voltage determines
the value of the conductivity far away from the ridge $\left|x\right|\gg W.$
However, the softness or the sharpness of the boundary is a function
of all the parameters somewhat equally.

\begin{figure*}[!t]
\begin{centering}
\includegraphics[scale=0.19]{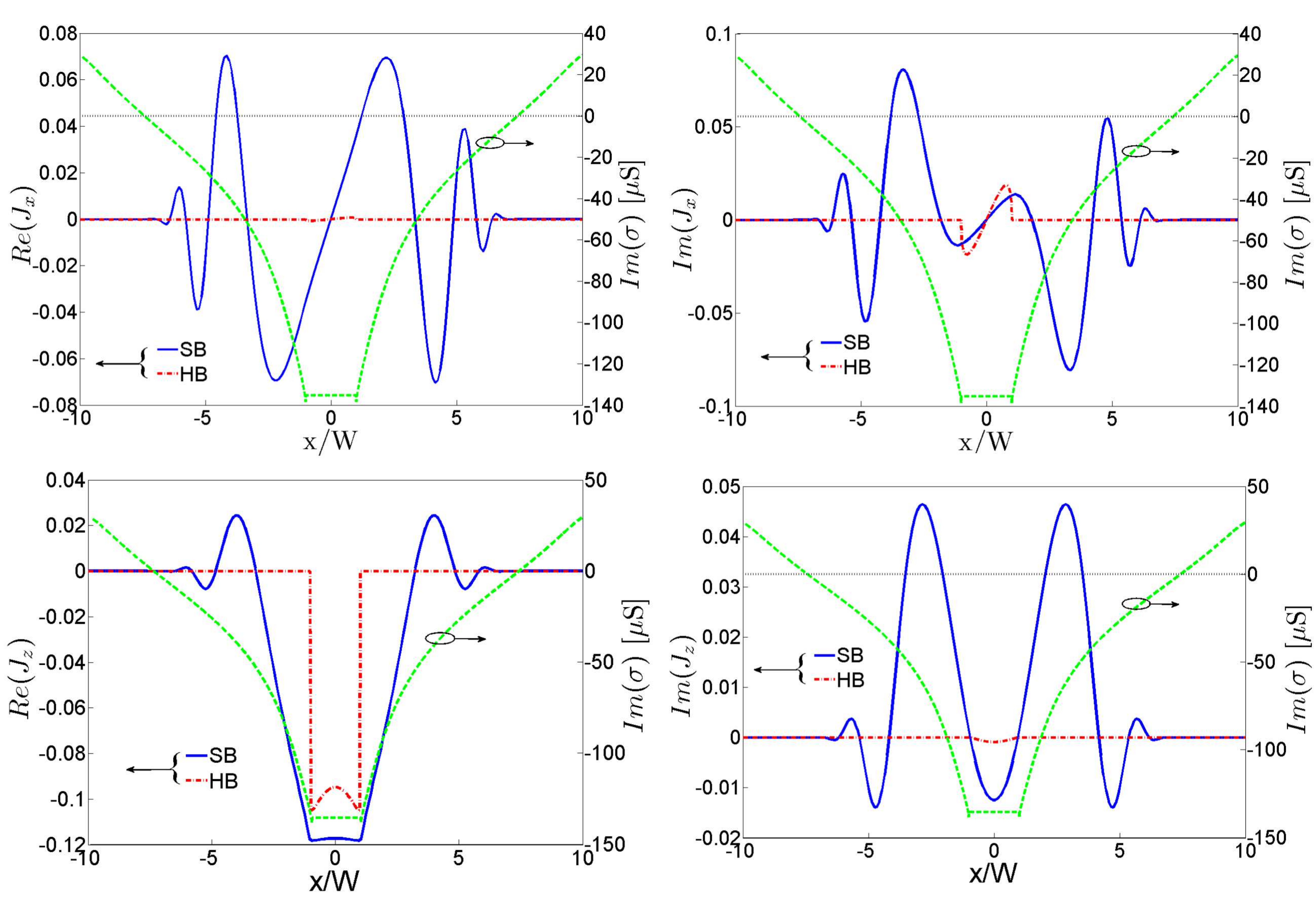}
\par\end{centering}

\caption{Longitudinal and transverse currents for the soft boundary (blue) and
the hard boundary (red) cases, and $\mathrm{Im}\left(\sigma\right)$
profile (green). Parameters are $f=30\: THz$, $W=25\: nm$, $a=25\: nm,$
$b=1\:\mu m,$ and $V_{0}=20\; V$.}
\end{figure*}

\begin{figure*}[!t]
\begin{centering}
\includegraphics[scale=0.4]{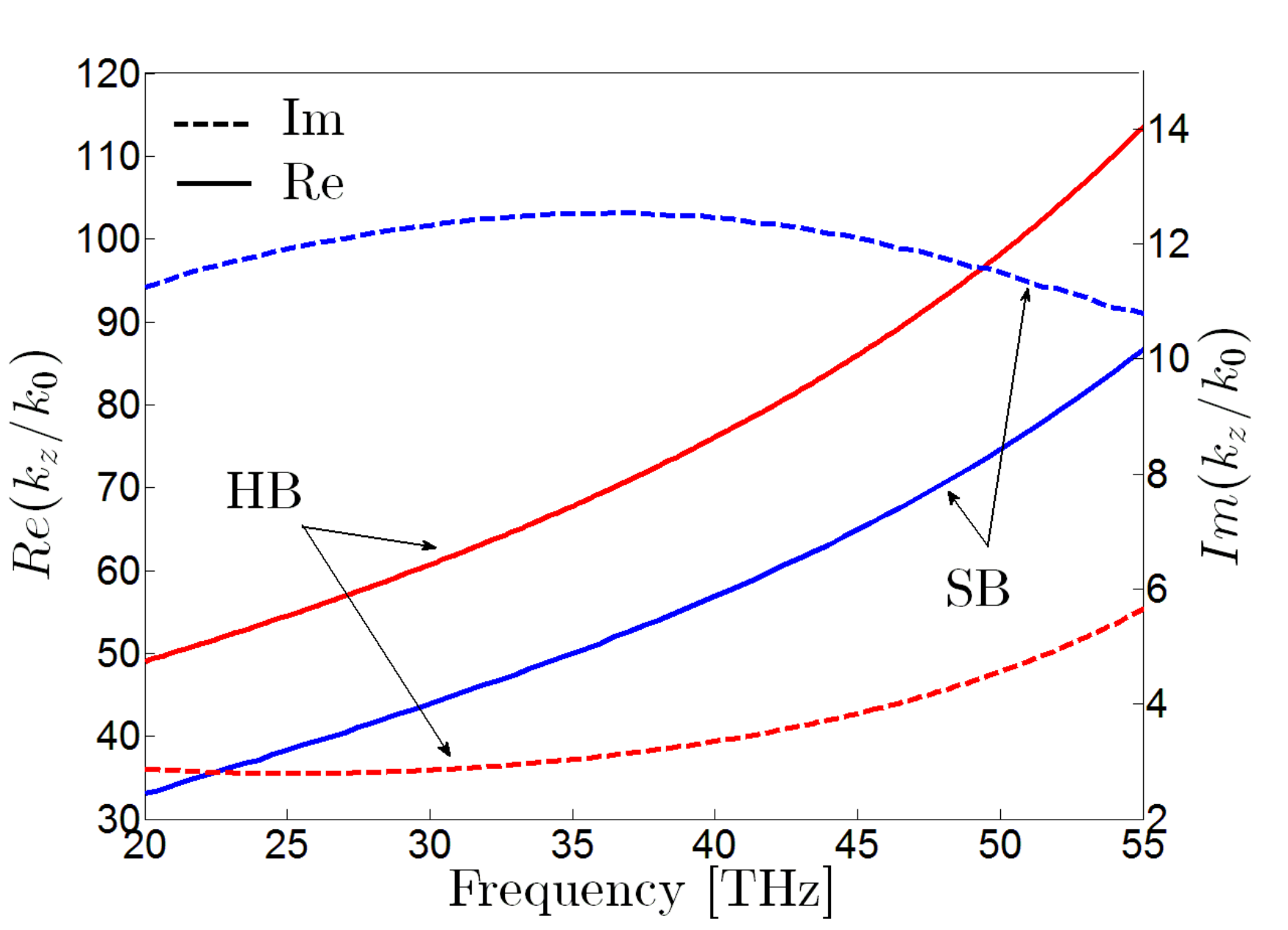}
\par\end{centering}

\caption{The dispersion curves of the fundamental mode for the soft boundary
 and the hard boundary  cases. Parameters are $W=25\: nm$,
$a=25\: nm,$ $b=1\:\mu m,$ and $V_{0}=20\; V$. }
\end{figure*}
 
\begin{figure*}[!t]
\begin{centering}
\includegraphics[scale=0.19]{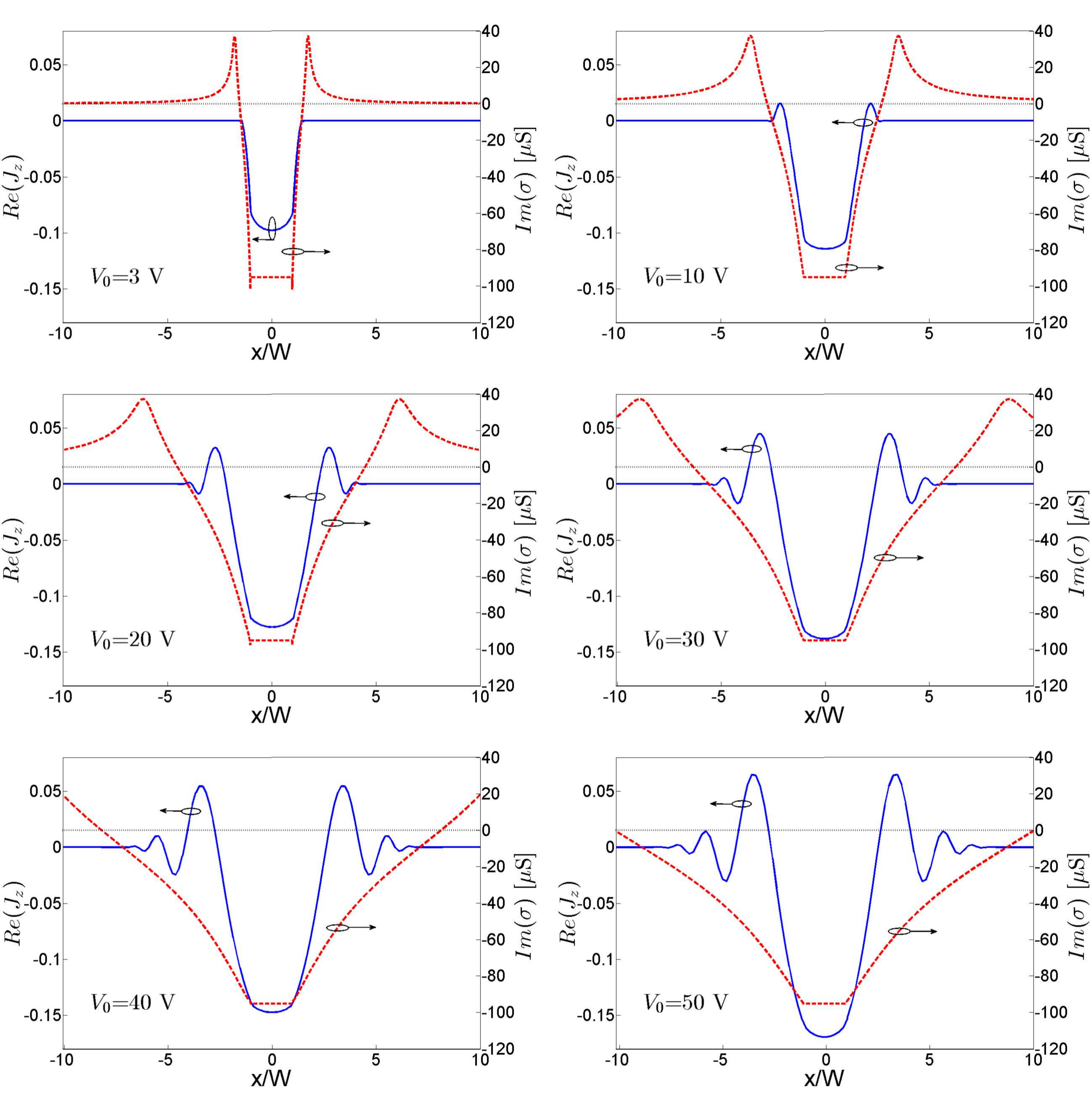}
\par\end{centering}

\caption{Real part of the longitudinal current for different values of $V_{0}$.
Other parameters are $f=40\:\mathrm{THz}$, $b=1\:\mu\mathrm{m}$
$W=25\:\mathrm{nm}$, and $a=1.25V_{0}\:\mathrm{nm}$. }
\end{figure*}

\begin{figure*}[!t]
\begin{centering}
\includegraphics[scale=0.19]{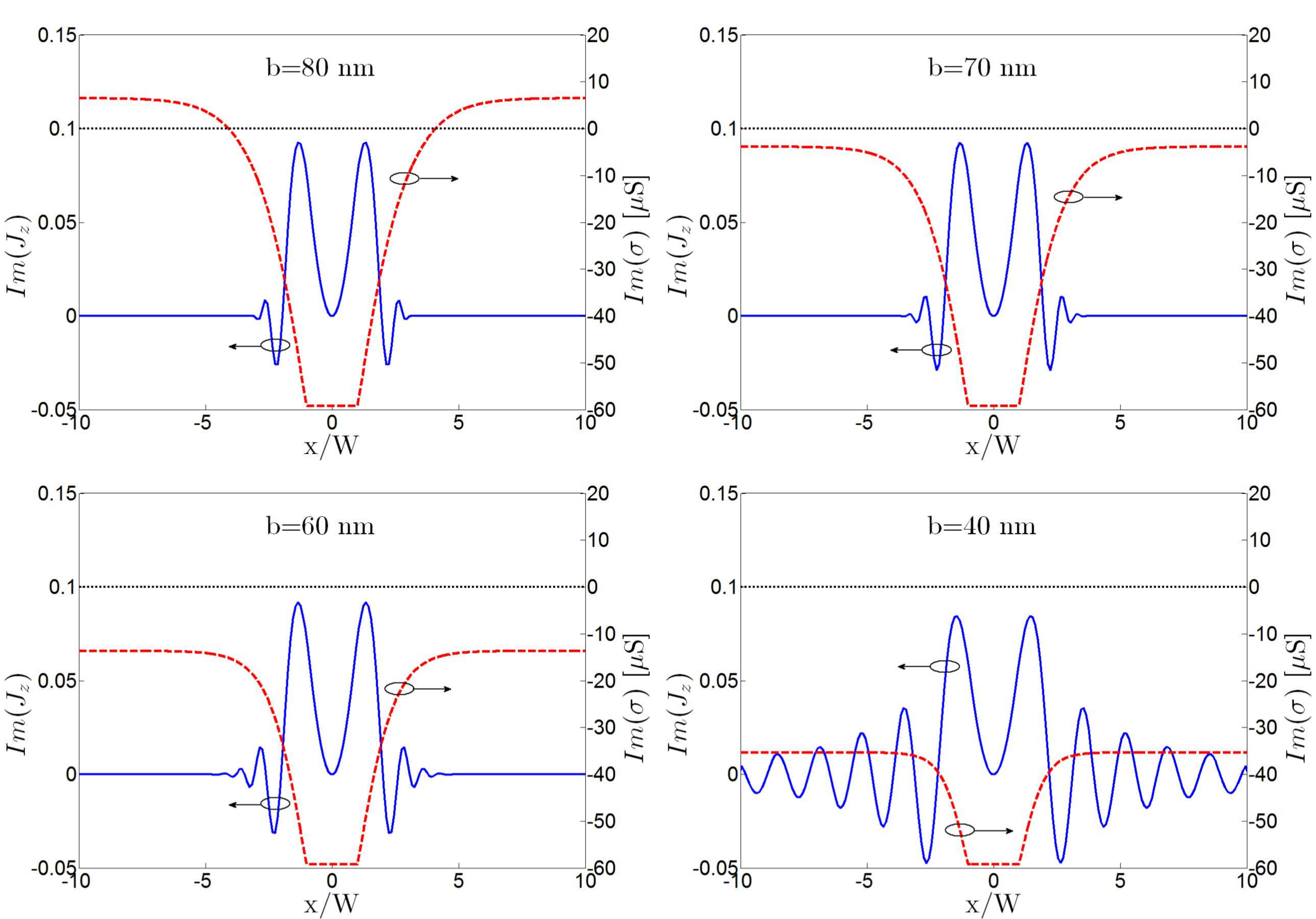}
\par\end{centering}

\caption{Imaginary part of the longitudinal current for different values of
$b$. Other parameters are $f=55\:\mathrm{THz}$, $V_{0}=20\:\mathrm{V}$
$W=25\:\mathrm{nm}$, and $a=25\:\mathrm{nm}$.}
\end{figure*}

Figure 4 shows the current distribution associated with the fundamental
SPP mode of figure 1. The conductivity and the current marked as SB
in figure 4 correspond to the geometry in figure 1 for $f=30\:\mathrm{THz}$,
$W=25\:\mathrm{nm}$, $a=25\:\mathrm{nm},$ $b=1\:\mathrm{\mu m},$
and $V_{0}=20\;\mathrm{V}$. The current which is noted as the hard
boundary current corresponds to a  graphene nanoribbon having width
of $50\,\mathrm{nm}$ and the same conductivity as the SB case for
$\left|x\right|<W$, and with $\sigma=0$ for $\left|x\right|>W$.
The dispersion curves associated with these currents are shown in
figure 5. The currents in figure 4 are normalized so that the 2-norm of
the eigencurrent vector (consisting of transverse and longitudinal
components) is unity, $\int\left(\left|J_{x}\left(x\right)\right|^{2}+\left|J_{z}\left(x\right)\right|^{2}\right)dx=1.$
Nonetheless, only the relative current component values are important
for our purposes. 

As figure 4 suggests, the current distribution for $\left|x\right|<W$
is similar for both soft and hard boundary cases (although $\mathrm{Re}\left(J_{x}\right)$
and $\mathrm{Im}\left(J_{z}\right)$ are much larger in the SB case)
and they both vanish as $\mathrm{Im}\left(\sigma\right)$ becomes
positive. However, the SB current has some oscillations near the two
boundaries. These oscillations resemble the field oscillations in
the cladding of an optical fiber with graded index cladding \cite{Kong}.
One of the consequences of this current spreading is that the mode
becomes more lossy since parts of the current flows in the region
with lower conductivity (soft boundaries). As an example, the propagation
constant for the SB and HB cases of figure 4 are $\beta_{z}/k_{0}=43.8-j12.3$
and $\beta_{z}/k_{0}=60.7-j2.8$, respectively. 

Figure 6 shows the effect of the boundary softness on the current distribution
($Re\left(J_{z}\right)$) of the fundamental mode, where $f=40\:\mathrm{THz}$,
$W=25\:\mathrm{nm}$, $b=1\:\mu\mathrm{m},$ and $V_{0}$ takes different
values. The parameter $a$ is set to be $a=1.25V_{0}\:\mathrm{nm}$
so that the conductivity values remains the same for $\left|x\right|<W$.
As figure 6 shows, the boundary becomes softer as $V_{0}$ increases
and the current oscillations increases (both in magnitude and number).

In figures 4 and 6 the currents vanish as $\mathrm{Im}\left(\sigma\right)$
becomes positive, which raises the question: is it necessary for $\mathrm{Im}\left(\sigma\right)$
to be positive away from the ridge to have a confined mode? To address this question, figure 7 shows $\mathrm{Im}\left(J_{z}\right)$
and conductivity distributions for different values of $b$; the other
parameters are $f=55\:\mathrm{THz}$, $W=25\:\mathrm{nm}$, $a=25\:\mathrm{nm},$
and $V_{0}=20\;\mathrm{V}.$ As figure 7 shows, $\mathrm{Im}\left(\sigma\right)$
changes sign for $b=80\:\mathrm{nm}$, but it remains negative everywhere
for $b=70,\,60,$ and 40 $\mathrm{nm}$. The currents remain
confined to the vicinity of the ridge region even for values of $b$
where $\mathrm{Im}\left(\sigma\right)$ remains negative everywhere.
However, as $b$ decreases the current spreads out further and the
mode becomes less confined. As a result, the important factor to
achieve good lateral mode confinement is that the ratio of (or the
difference between) $\mathrm{Im}\left(\sigma\right)$ above and away
from the ridge should be large.

\section{Conclusion}

The conductivity and the current distributions were studied for an
infinite graphene sheet over a ridge-perturbed ground plane. It was
shown analytically that a channel with soft boundaries will be formed
above the ridge to guide SPPs provided that the parameters are adjusted
properly. It was also shown that the width of the channel is more
sensitive to the bias voltage than the geometric ridge parameters.
It was observed that the SPP can be kept confined to the vicinity
of the ridge even if the formed channel does not have a finite width
(i.e., that $\mathrm{Im}\left(\sigma\right)$ is negative everywhere)
provided that the channel boundaries have sharp enough slopes. Since the width of the formed channel can be controlled by both frequency and the bias voltage, the spatial location of the current concentration (and its associated field) for a surface mode can be controlled. This can be useful in switching or frequency demultiplexing applications.

\section*{References}

\bibliography{IOP}

\end{document}